\documentclass{segabs}

\usepackage{amsmath}
\usepackage{booktabs}
\usepackage{placeins}

\usepackage[outdir=./]{epstopdf}

\begin{document}

\title{Application of Co-Kriging and Ordinary Kriging for Selecting Additional Well Locations}

\renewcommand{\thefootnote}{\fnsymbol{footnote}} 
\renewcommand{\figdir}{Fig} 

\author{Rong Lu\footnotemark[1], Colorado School of Mines}

\footer{Example}
\lefthead{Lu}
\righthead{Using co-kriging and ordinary kriging for finding new well locations}

\maketitle

\begin{abstract}
  Well performance index (WPI), which is an indicator on how much producing potential a well has, is proposed for Cana Field using the available information from the completion database. I used ordinary kriging and co-kriging to create interpolation maps for WPI across the region. The interpolation results can be used to predict WPI values for locations that have not gone through drilling programs, thus guiding operator to find the next drilling locations. Different kriging models' performance are compared using cross-validation. It is shown co-kriging with clean fluid volume has the best performance. Recommendations are given regarding new well locations. 
\end{abstract}

\section{Introduction}

Unconventional oil and gas resources development has gained much more attention since the last decade, due to the advancement in hydraulic fracturing (HF, or ``frac'') technology. In order to develop shale gas reservoirs, which have extremely low permeability, HF has to be applied. In the process fluids and solids under high pressure are pumped into the formation to break the rock. As fractures are created, more reservoir contact are obtained and the shale gas would flow through the fractures to the wellbore. One question the industry are interested in is, where to drill/frac the wells in unconventional shale plays.

In this work, a well performance index (WPI) is used to evaluate the production potential for existing wells (``samples''), which leverages the initial pressure and production rates information. Then geostatistical methods, including ordinary kriging and co-kriging, are used to predict producing potentials for un-drilled regions (``unsampled'' locations). My goal is to compare the performance of different estimators:
\begin{enumerate}
    \item ordinary kriging (\textbf{OK}) on variable of interests
    \item co-kriging (\textbf{CK}) using secondary variables with different levels of correlation
\end{enumerate}
using cross-validation, and then propose new drilling location candidates for Cana Field from the found best estimator. 

The Cana Field dataset \cite[]{lu2014investigation} is a well completion database. It contains more than 400 shale gas wells drilled in Cana Woodford Shale in Oklahoma; for each well, there are completion and HF job parameters, and initial production data up to the first 90 days.

\section*{Theory \& Method}

In this section how to find new locations for HF wells is discussed. From the literature \cite[]{Konstantin}, a well performance index (WPI) is used to estimate how much production potential a well has, using the initial 90 days' production and pressure data:
\begin{equation}
\text{WPI}=\sum_{i=1}^{90}\text{dailyProdRate}_i \times \text{dailyPressure}_i
\end{equation}
    It is assumed in this work that pressure is constant over the first 90 days and that fracture pressure can be used for estimation purposes, then the equation becomes (the hat indicates it is an estimator for the true WPI):
\begin{multline}
\label{eqn:wpiHat}
\widehat{\text{WPI}}=\text{avgProdRateFirst90Days(MCFE/day)} \times 90 \times \\ 
\text{fracGradient} \times \text{TVD}
\end{multline}
    Now all the information needed is available from the dataset. A visualization of the spatial distributions and relative values of $\widehat{\text{WPI}}$ is shown in Figure~\ref{fig:wpiMap}. The approach for prediction, known as ``kriging'', is essentially a linear estimator using the known information. The core idea is to assign weights $w_i$ to each known data point $z(x_i)$ located at $x_i$, and by applying a linear summation the property's value at unsampled location $x_0$ is obtained:
\begin{equation}
\widehat{Z}(x_0) = 
        \begin{bmatrix}
            w_1       & w_2 & \dots & w_n 
        \end{bmatrix} 
        \cdot
        \begin{bmatrix}
            z_1 \\ z_2 \\ \vdots \\ z_n 
        \end{bmatrix} = \sum_{i=1}^{n} w_i(x_0) \times Z(x_i)
\end{equation}
    Weights, $w_i$, are determined by minimizing variance of estimation \cite[]{gstat}. The performance measure of this linear estimator will be discussed later. In OK, only the variable of interests (primary variable or primary data) is studied and the weights will add up to unity. CK also takes advantage of secondary-data, which have spatial correlations with primary data. Then the covariance between two or more regionalized variables is leveraged. 

\plot{wpiMap}{width=\columnwidth}
{
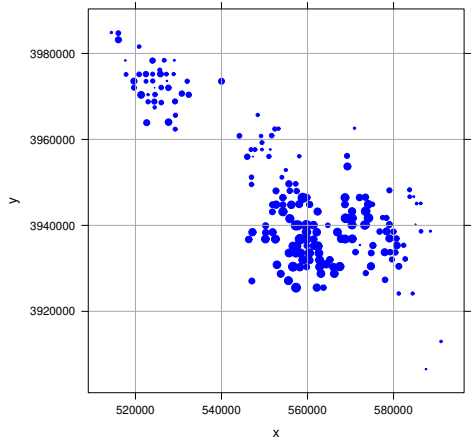
This shows a bubble plot for WPI samples. Larger points indicate larger WPI values. $X$ and $Y$ axes are for easting and northing (units in meters), respectively.
}

\subsection{Assumptions and data pre-processing}

Besides using the approximated WPI (Equation~\ref{eqn:wpiHat}), only co-located primary/secondary data is included in this study, which ensures a linear model of coregionalization (will be addressed later). Since the original values of variables of interests are huge, both primary and secondary variables are log10 transformed.

\subsection{Secondary variables}

Clean fluid volume and proppant volume are chosen to be the secondary data respectively. Their corresponding correlations with primary data is shown in Figure~\ref{fig:fluidVsWPI,propVsWPI}. It can be seen fluid volume has higher correlation with WPI than proppant volume. 

\multiplot{2}{fluidVsWPI,propVsWPI}{width=\columnwidth}
{
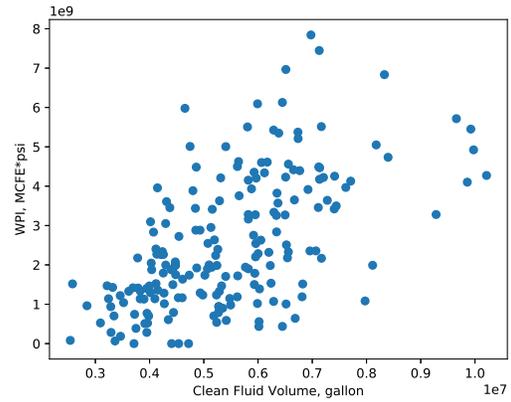
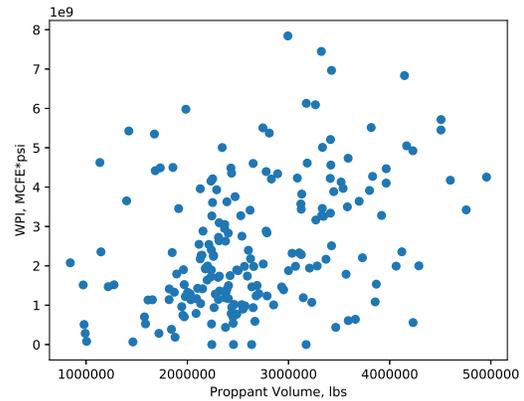
Scatter plots between primary and secondary variables. (a) WPI and fluid volume have correlation coefficient of 0.58; (b) WPI and proppant volume have correlation coefficient of 0.37.
}

\subsection{Toolset}
Stanford Geostatistical Modeling Software (SGeMS) is used for the purpose of exploratory data analysis (EDA). For the final results generation and presentation, they are completed by two \texttt{R} packages: \texttt{gstat} \cite[]{gstat} and \texttt{sp} \cite[]{sp}. These packages provides:
      \begin{itemize}
          \item OK and CK algorithms
          \item API for cross validation
          \item API for fitting linear models of coregionalization
      \end{itemize}

\subsection{Strategy for verification}

In this work $k$-fold cross validation is conducted for comparing the performance of different estimators. Figure~\ref{fig:cvDemo} shows an illustration when the number of folds, $k$, equals 5. Since the number of WPI samples is 190 (not huge), leave-$one$-out (LOO) is performed: each run I use 189 data points to ``krige'' the value at the location of the 190th point, and then calculate the residual by comparing kriging result with the true value. Two performance measures are adopted:
\begin{enumerate}
    \item Mean error:
        \begin{equation}
        \label{meanErr}
        \text{ME}=\dfrac{1}{n} \sum_{i=1}^{n} (y_i-\widehat{y_i})
        \end{equation}
    \item Root-mean-square error:
        \begin{equation}
        \label{rmse}
        \text{RMSE}= \sqrt{\dfrac{1}{n} \sum_{i=1}^{n} (y_i-\widehat{y_i})^2}
        \end{equation}
\end{enumerate}

\plot{cvDemo}{width=\columnwidth}
{
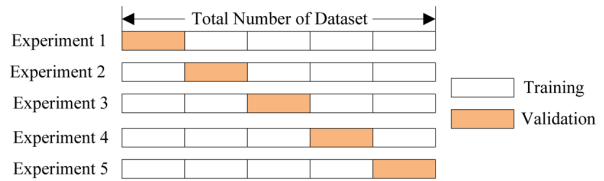
This illustrates the case of 5-fold cross-validation. The original sample is randomly partitioned into 5 equal sized subsamples. 4 of the 5 subsamples will be used to train the model, being tested against the held-out 1 subsample. The process repeats 5 times.
}

\FloatBarrier

\section*{Results \& Verification}
\subsection{Ordinary kriging}

The prediction map generated by OK is shown in Figure~\ref{fig:okVariogram,wpiOkMap}

\multiplot{2}{okVariogram,wpiOkMap}{width=\columnwidth}
{
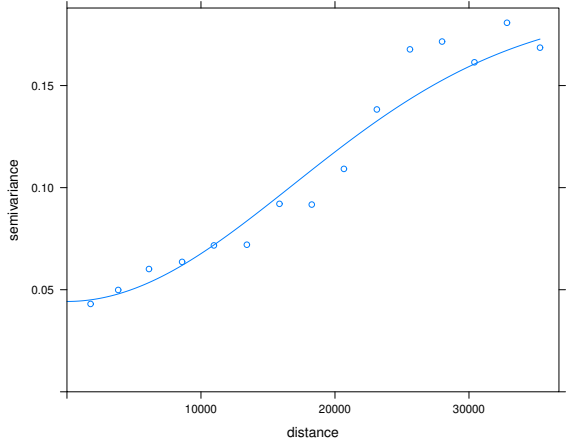
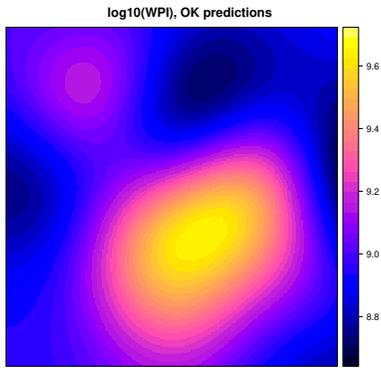
(a) Variogram modeling of the primary variable WPI; (b) WPI map generate by OK.
}

\FloatBarrier

\subsection{Co-kriging with fluid volume}

In CK, the spatial covariance between primary and secondary data is represented by a cross variogram, as presented in Figure~\ref{fig:ckFluidVarios,ckFluid_WpiMap} along with the output WPI map.

\multiplot{2}{ckFluidVarios,ckFluid_WpiMap}{width=\columnwidth}
{
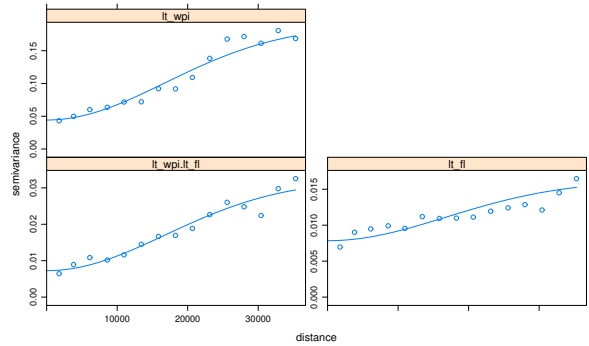
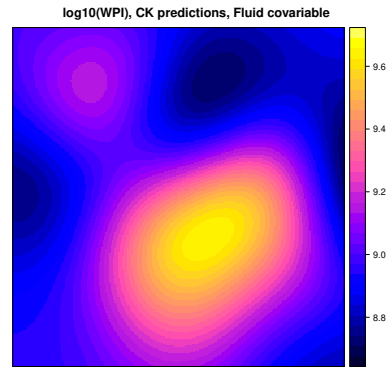
(a) Fitted direct and cross variograms between log10(WPI) and log10(Fluid). ``lt'' and ``fl'' stands for log10 and fluid volume respectively; (b) WPI map generated by CK with fluid volume.
}

\subsection{Co-kriging with proppant volume}

CK results using proppant volume as secondary data is shown in Figure~\ref{fig:ckPropVarios,ckProp_WpiMap}. Note the proppant volumes' experimental variogram has a fluctuation behavior naturally but the chosen variogram model smooths it out. This is also to ensure a linear model of coregionalization (will be addressed later).

\multiplot{2}{ckPropVarios,ckProp_WpiMap}{width=\columnwidth}
{
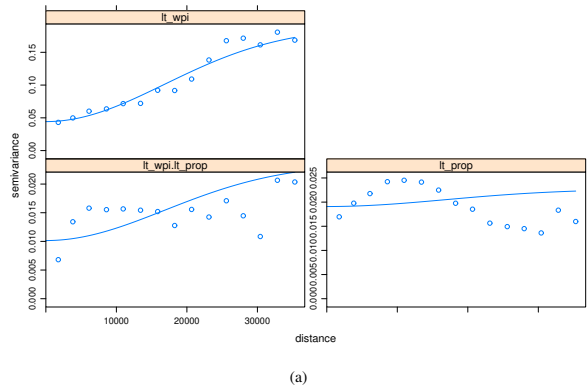
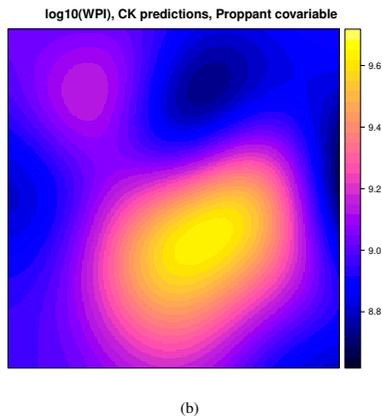
(a) Fitted direct and cross variograms between log10(WPI) and log10(Proppant). ``lt'' and ``prop'' stands for log10 and proppant volume respectively; (b) WPI map generated by CK with proppant volume.
}

\FloatBarrier

\subsection{Estimator comparisons}
Using the performance measures discussed before (Equation~\ref{meanErr} and Equation~\ref{rmse}), the comparisons are summarized in Table~\ref{tbl:modelCompar}. It can be seen CK results are more accurate than OK result, though CK takes much longer time to run. CK with fluid is chosen as the best model since its modeled variogram honors the experimental variogram better than CK with proppant did (refer to Figure~\ref{fig:ckFluidVarios} and Figure~\ref{fig:ckPropVarios}).

\tabl{modelCompar}{
Comparison of the Three Interpolations
}{
  \begin{center}
\begin{tabular}{l|cccc}
\toprule
     & Mean Error & RMSE & Running Time \\ \hline
     OK & -0.0014 & 0.249& 1.86 sec \\
     CK with fluid & -0.0007 & 0.232  & 6.74 sec \\ 
     CK with proppant & -0.0009 & 0.229 & 6.67 sec \\ 
\bottomrule
\end{tabular}
  \end{center}
}

\section{Discussion - Conclusion}

One of the benefits CK provides is that it might give better predictions when the main attribute of interest (primary variable; WPI in this case) is sparse, but related secondary information is abundant. The raw data from Cana dataset contains more fluid volume (also proppant volume) than WPI, motivating me to take the CK approach. For modeling, one needs to fit both the direct and cross-variograms simultaneously, the results of which must produce a positive definite system. Easiest way to ensure this is to fit a linear model of coregionalization: all the variograms have same shape and range, with different partial sills and nuggets. As presented in Figure~\ref{fig:fullDataCrossVarios}, when including all the secondary data (fluid volume), fluid volume has naturally a shorter range than the primary variable. Thus finally only co-located data points are included for modeling which happen to make secondary variables have approximately the same range as the primary variable. 

\plot{fullDataCrossVarios}{width=\columnwidth}
{
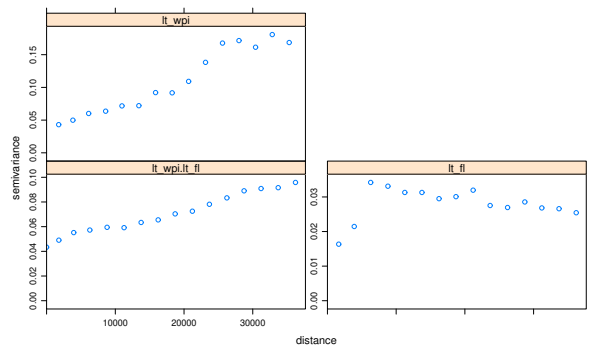
Experimental variograms when including all the available secondary data. ``lt'' and ``fl'' stands for log10 and fluid volume respectively.
}

In summary, it has been shown that co-kriging performs better than ordinary kriging on the Cana dataset. The CK with fluid volume model provides the best predictive power regarding which regions have higher producing potentials, from where new drilling locations are proposed (Figure~\ref{fig:drillDirecPDF}).

\plot{drillDirecPDF}{width=\columnwidth}
{
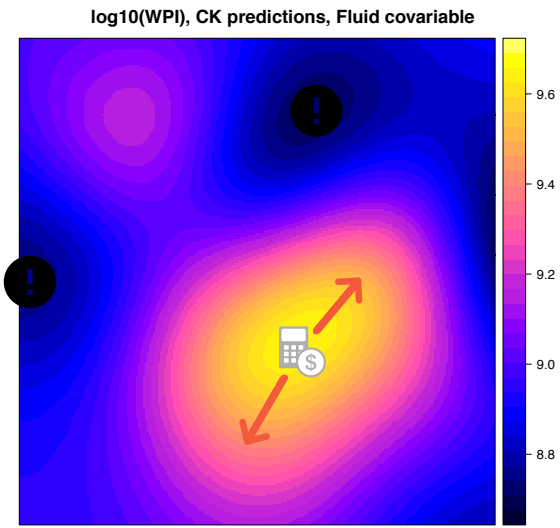
New wells are recommended to be located at the ``dollar sign'' area, whereas the zones showing warnings have lower producing potentials.
}

\bibliographystyle{seg}  
\bibliography{example}

\end{document}